\documentclass[a4paper,fleqn,usenatbib,useAMS]{mnras}

\usepackage{graphicx}	
\usepackage{amsmath}	
\usepackage{amssymb}	
\usepackage{multicol}        
\usepackage{bm}		
\usepackage{pdflscape}	
\usepackage{natbib}
\usepackage{times}
\usepackage{calc}
\usepackage{rotating}
\usepackage{lscape}
\bibpunct{(}{)}{;}{a}{}{,} 
\usepackage{longtable}
\usepackage{enumitem} 
\usepackage{hyperref}
\usepackage{pifont}
\usepackage{tikz}

\def\oversim#1#2{\lower0.5pt\vbox{\baselineskip0pt \lineskip-0.5pt
     \ialign{$\mathsurround0pt #1\hfil##\hfil$\crcr#2\crcr\sim\crcr}}}

\usepackage[T1]{fontenc}
\usepackage{ae,aecompl}

\usepackage{mathptmx}

\def\aap {{A\&A}}

\def\apjl {{ApJ}}

\def\apjs {{ApJs}}

\title[Submm analysis of Frosty Leo]{ A Submillimeter Polarization Analysis of Frosty Leo}

\author[L. Sabin, Q. Zhang, R. V\'azquez and W. Steffen] 
{L. Sabin$^{1}$\thanks{E-mail:lsabin@astro.unam.mx (LS)}, Q. Zhang$^{2}$, R. V\'azquez$^{1}$ and W. Steffen$^{1}$ \\  
$^{1}$Instituto de Astronom\'{\i}a, Universidad Nacional Aut\'onoma de M\'exico, Apdo. Postal 877, 22800 Ensenada, B.C., Mexico\\
$^{2}$Harvard-Smithsonian Center for Astrophysics, 60 Garden Street, Cambridge, MA 02138, USA}

\date{Accepted 2019 January 10. Received 2019 January 10; in original form 2018 June 14}
\pubyear{2017}

\begin{document}
\label{firstpage}
\pagerange{\pageref{firstpage}--\pageref{lastpage}}
\maketitle


\begin{abstract}

We present a polarimetric investigation of the protoplanetary nebula Frosty Leo performed with the Submillimeter Array. We were able to detect, in the low continuum level (peak at 14.4 mJy beam$^{-1}$), a marginal polarization at $\sim2.6\sigma$. 
The molecular line investigation based on the CO $J=3\rightarrow2$ emission shows a peak 
emission of 68.1\,Jy\,beam$^{-1}$\,km\,s$^{-1}$ and the polarization detection in this CO line is also marginal, with a peak at $\sim3.8\sigma$. In both cases, it was therefore not possible to use the electric vector maps ($\bmath{E}$-field) to accurately trace the magnetic field ($\bmath{B}$-field) within the PPN.
 The spatio-kinematic modelling realised with the different velocity channel maps indicates three main structures: a distorted torus accompanied by a bipolar outflow or jet aligned with its axis and a 
flattened spherical ``cap''. The comparison of the CO polarization segments with our model suggests that the polarized emission probably arises in the first two components. 
 
\end{abstract}


\section{Introduction}\label{intro}
 
Many observational efforts toward polarimetric measurements, have been performed to understand the role of magnetic fields in (the geometry of) evolved intermediate 
mass stars such as proto-planetary nebulae (PPNe) and planetary nebulae (PNe) (\citealt{Rodriguez2017,Gomez2009,Vlemmings2008,Bains2004}). 
Following our previous works aiming at detecting and mapping magnetic fields via the linear polarization of both the dust continuum and molecular line 
emission\footnote{A description of the methods can be seen in \citet{Sabin2014}}(see \citealt{Sabin2007,Sabin2014,Sabin2015}), we present in this article a 
polarization analysis of the PPN Frosty Leo (IRAS\,09371+1212) performed with the Submillimeter Array (SMA).

Classified as a bipolar post-AGB star or a PPN by \citet{Forveille1987} and \citet{Rouan1988}, Frosty Leo was named after the presence of a sharp 
peak at 60$\mu$m indicative of cold icy grains. \citet{Forveille1987} also showed the presence of a strong CO ($J=1\rightarrow0$) emission in the envelope with 
a $V_{\rm LSR}= -10$\,km\,s$^{-1}$ and an expansion velocity of 25\,km\,s$^{-1}$ indicative of an outflow. Located at a distance estimated between 1 and 4\,kpc 
by \citep{Mauron1989} and $3.08\pm0.71$\,kpc by \citep{Vickers2015}, Frosty Leo has been the subject of various morphological studies.  \citet{Kwok1993} and 
\citet{Langill1994} noticed that the bipolar nebula is surrounded by a nearly spherical envelope of material approximately of 30\,arcsec in diameter. \citet{Beuzit1994} 
identified the disc-like structure of the equatorial plane and \citet{Roddier1995} showed the presence of a companion to the K7III type Central Star (CS). In addition, 
they showed that the central region of the PPN is not heavily obscured and that the CS lies in a $\sim10^3$\,au thick disc of $\sim4\times10^3$\,au in diameter 
(they assumed a distance of 1.27\,kpc) for their calculations. The lobes are found to extend up to 15$\times$10$^{3}$\,au. The authors found that the collimated 
ejected material from the centre of the nebula seemed to be distributed within a disrupted cone. Subsequently, \citet{Sahai2000}, using optical HST data, reported 
a complex bipolar nebula displaying multiple jets close to the equatorial plane and two bright ansae, one on 
each side of an edge-on disc. The $^{12}$CO molecular line 
observations realized by \citet[hereafter CC05]{Carrizo2005} reveal a rather compact configuration of which two main attributes are a ring like structure and high 
speed jets showing an expansion velocity $\sim75$\,km\,s$^{-1}$. The latter are also tracing the optical jets.

The main objectives of this investigation are to determine  (i) if a polarization signature is present in the circumstellar envelope of Frosty Leo, (ii) if/how the 
polarisation pattern could be associated to the occurrence of a magnetic field and (iii) if/how the morphological features mentioned above can be, entirely or 
partially, linked to these polarization patterns and a possible action of the magnetic field. The article is organized as follows: in Section \ref{obs} we describe the 
observations and data reduction process, in Section \ref{continuum} and Section \ref{line} we present the continuum and line polarization results respectively. 
The modelling of the CO emission is shown in Section \ref{model}. Finally, the concluding remarks are presented in Section \ref{conclusion}.

\section[]{Observations}\label{obs}

\begin{figure}
\includegraphics[height=7.5cm]{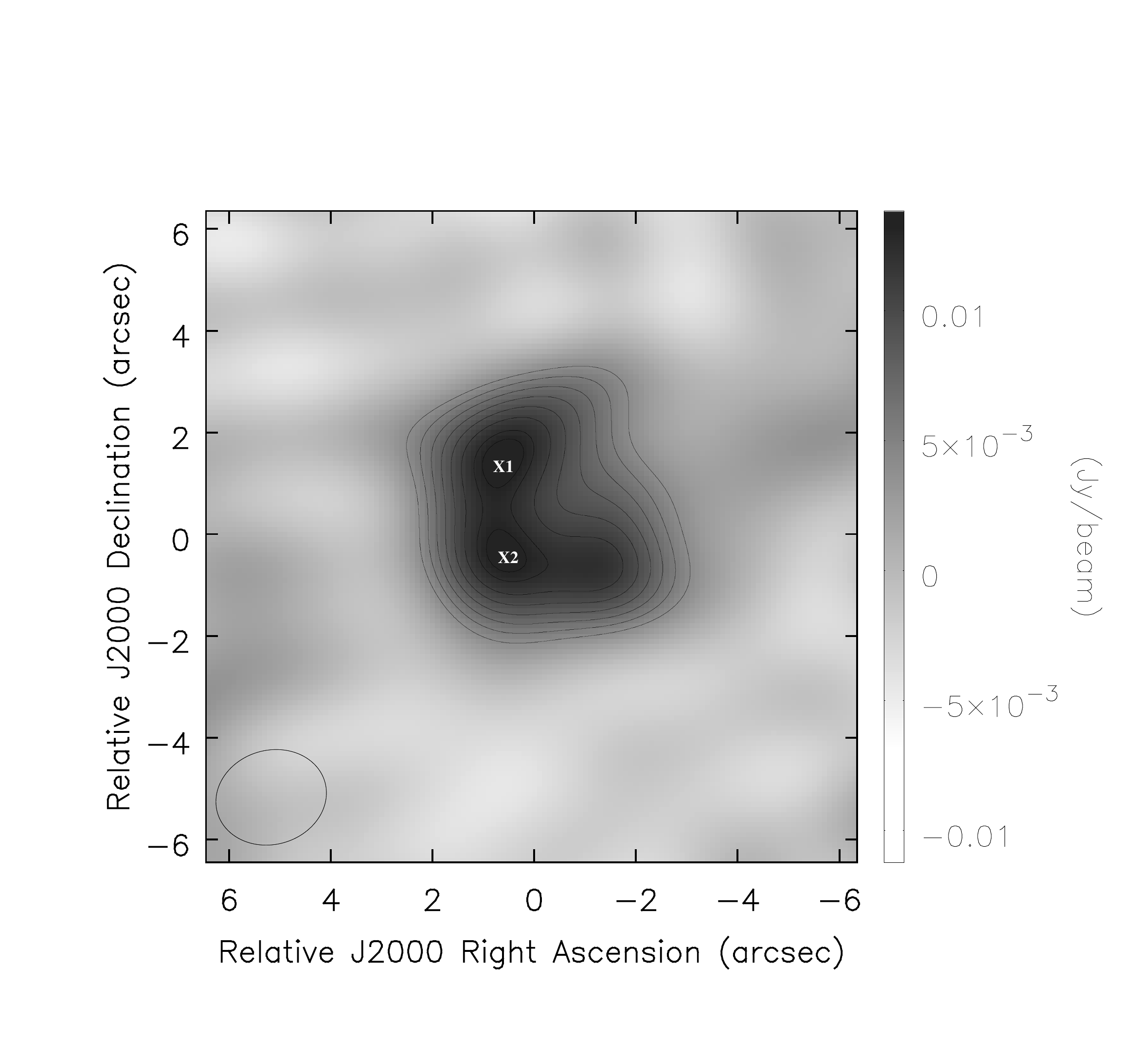}
\caption{Continuum emission at 345\,GHz of Frosty Leo. The solid white contours are drawn in steps of 10\% starting at 30\% of the peak emission and the wedge 
on the right indicates the continuum flux in Jy\,beam$^{-1}$.  The peak fluxes  corresponding to the two bright lobes are marked as ``x1'' with a flux of 
14.2\,mJy\,beam$^{-1}$ and ``x2'' with a flux of 14.4\,mJy\,beam$^{-1}$. The negative contours are shown as dash lines. Coordinates of the origin are $\alpha({\rm J 2000})=09^{\rm h}39^{\rm m}53\fs959$, 
$\delta({\rm J 2000})=+11\degr58\arcmin52\farcs60$ . North is up and east is left.}
\label{cont}
\end{figure} 

\begin{figure}
\begin{center}
\includegraphics[height=7.5cm]{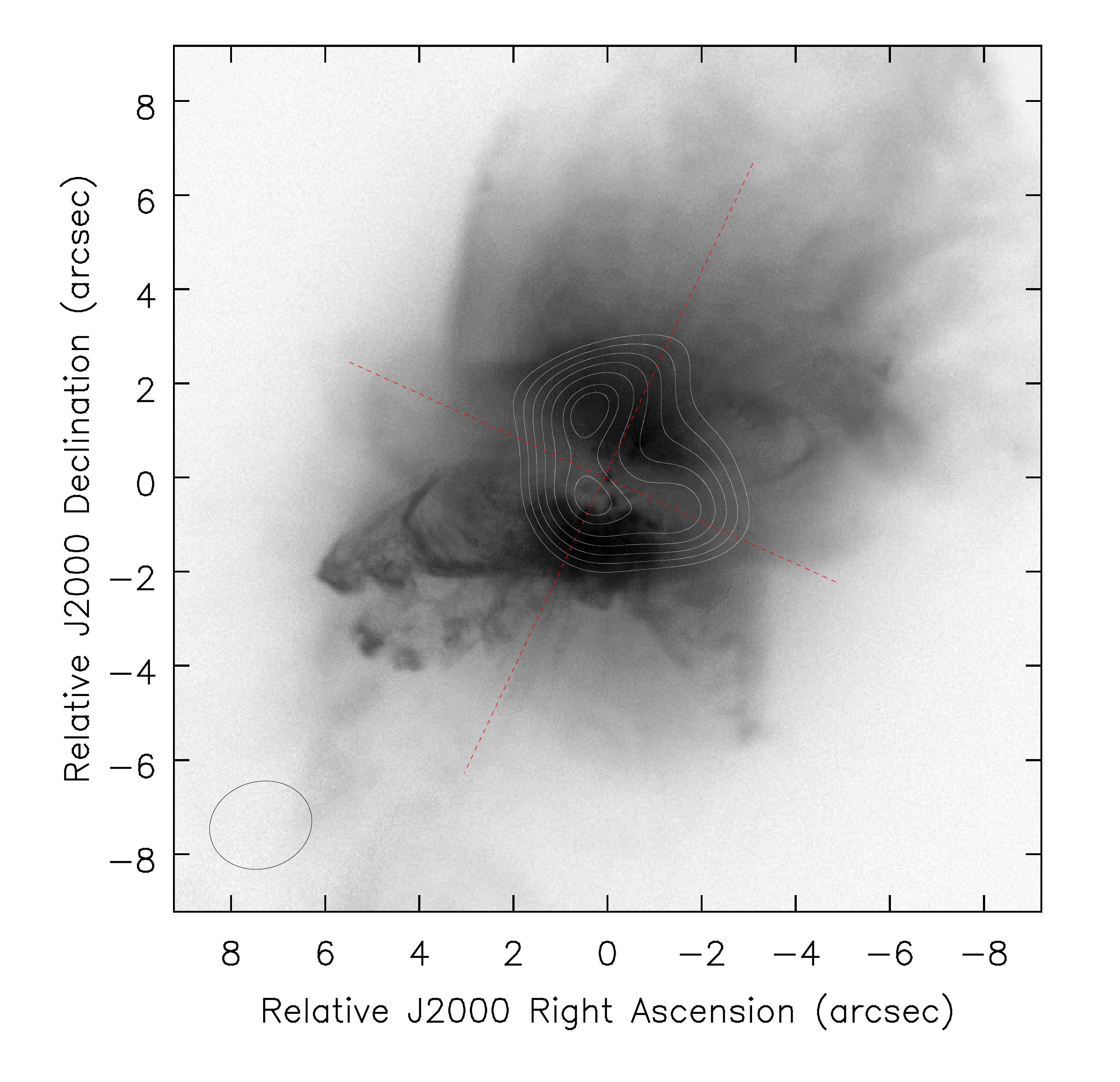}
\caption{HST composite images of Frosty Leo with the continuum emission contours superimposed. Contours are drawn in steps of 10\% starting at 30\% of the 
peak emission. The dotted lines indicate the PPN symmetry axis. }
\label{hst}
\end{center}
\end{figure} 

The polarimetric observations were performed with the Submillimeter Array \citep[SMA\footnote{The Submillimeter Array is a joint project between the Smithsonian 
Astrophysical Observatory and the Academia Sinica Institute of Astronomy and Astrophysics, and is  funded  by  the  Smithsonian  Institution  and  the Academia Sinica.};]
[]{Ho2004,Rao2005} on 2016 January 28. The observations, which lasted $\sim12$\,h including calibration, were performed using the compact configuration. 
However out of eight available antennas only six were operating due to technical issues. The weather conditions were excellent during the run with $\tau\simeq$0.03 
at 225\,GHz and relatively stable phases during the observations. The total frequency coverage was from $\sim$334\,GHz to 346\,GHz with a gap between 336 and 
342\,GHz marking the division between the lower and upper side bands (LSB and USB respectively). It was therefore possible to target the usually strong 
$^{12}$CO $J=3\rightarrow2$ line at rest frequency 345.796\,GHz. This setting is suitable to make some comparison with the CO findings of CC05 for example. 
The quasar J0854+201 was used as the gain calibrator and 3C84 as a bandpass and polarization calibrator. The data reduction process, which involves flux, gain 
and bandpass calibration, was performed with the software package {\sc mir}\footnote{Available at https://www.cfa.harvard.edu/$\sim$cqi/mircook.html} and then 
exported to the Multichannel Image Reconstruction, Image Analysis and Display software {\sc miriad} (\citealt{Wright1993,Sault2011}) for polarization calibration 
and imaging. The data were also corrected for polarization leakage; we found consistent leakage terms of a few per cent in the LSB and USB.
\begin{figure*}
\begin{center}
\includegraphics[height=7cm]{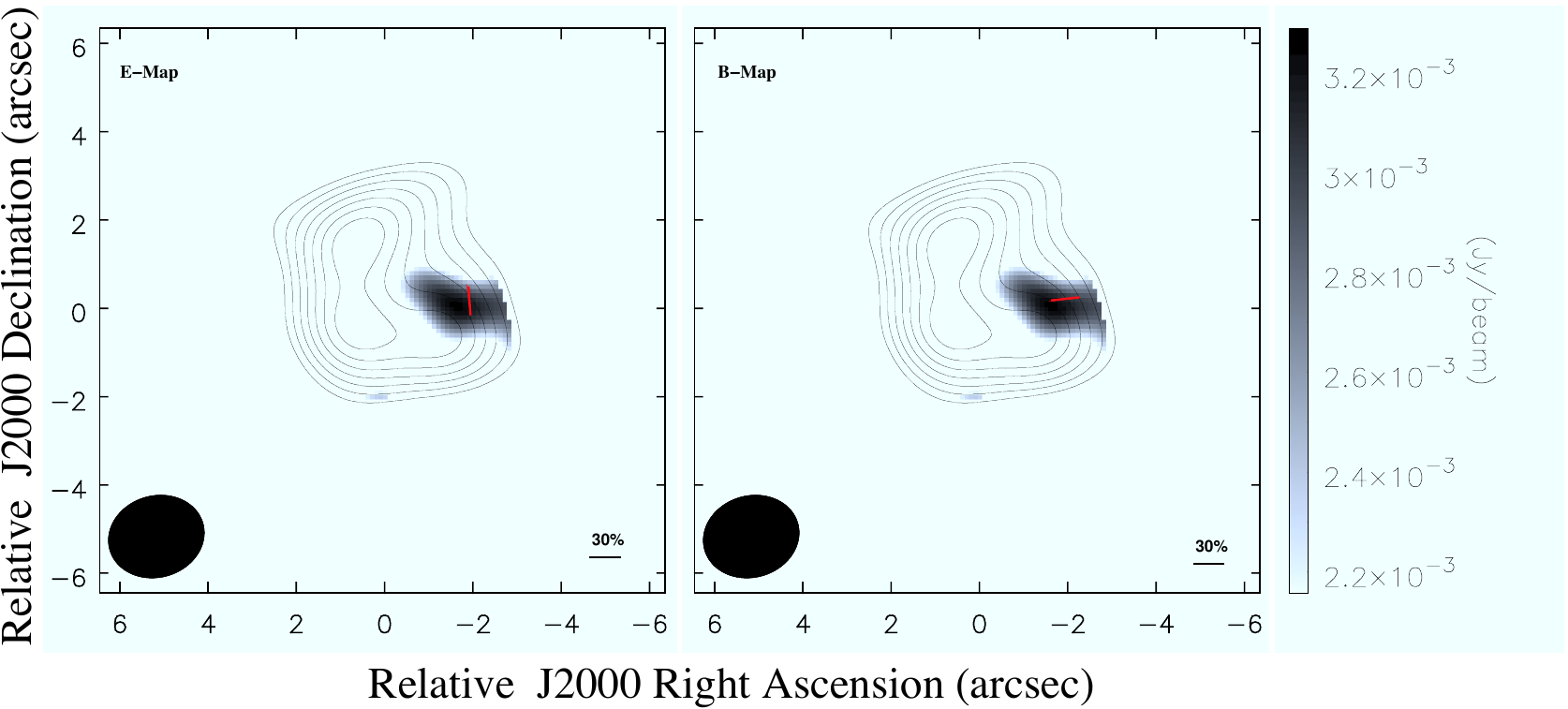}
\caption{Left: Polarization map indicating the location of the polarized emission in grey scale and the $\bmath{E}$-field is superimposed as the red segment. Only one segment 
is plotted as the polarized emission is not spatially resolved {\it(see text)}. The contours indicate the continuum intensity as described in Fig.~\ref{cont}. Right: The possible 
``magnetic field map'' derived from the $\bmath{E}$-segments rotated by $90\degr$. In both cases the wedge on the right indicates the polarized flux in Jy Beam$^{-1}$. The beam is 
drawn in the left-bottom corner of each panel. The cut for the polarization emission is 2.2 mJy/beam in both maps. The scale bar in the right-bottom corner represents the percentage of polarization (see text).}
\label{sma}
\end{center}
\end{figure*} 

\begin{figure}
\hspace{-1cm}
\includegraphics[height=7cm]{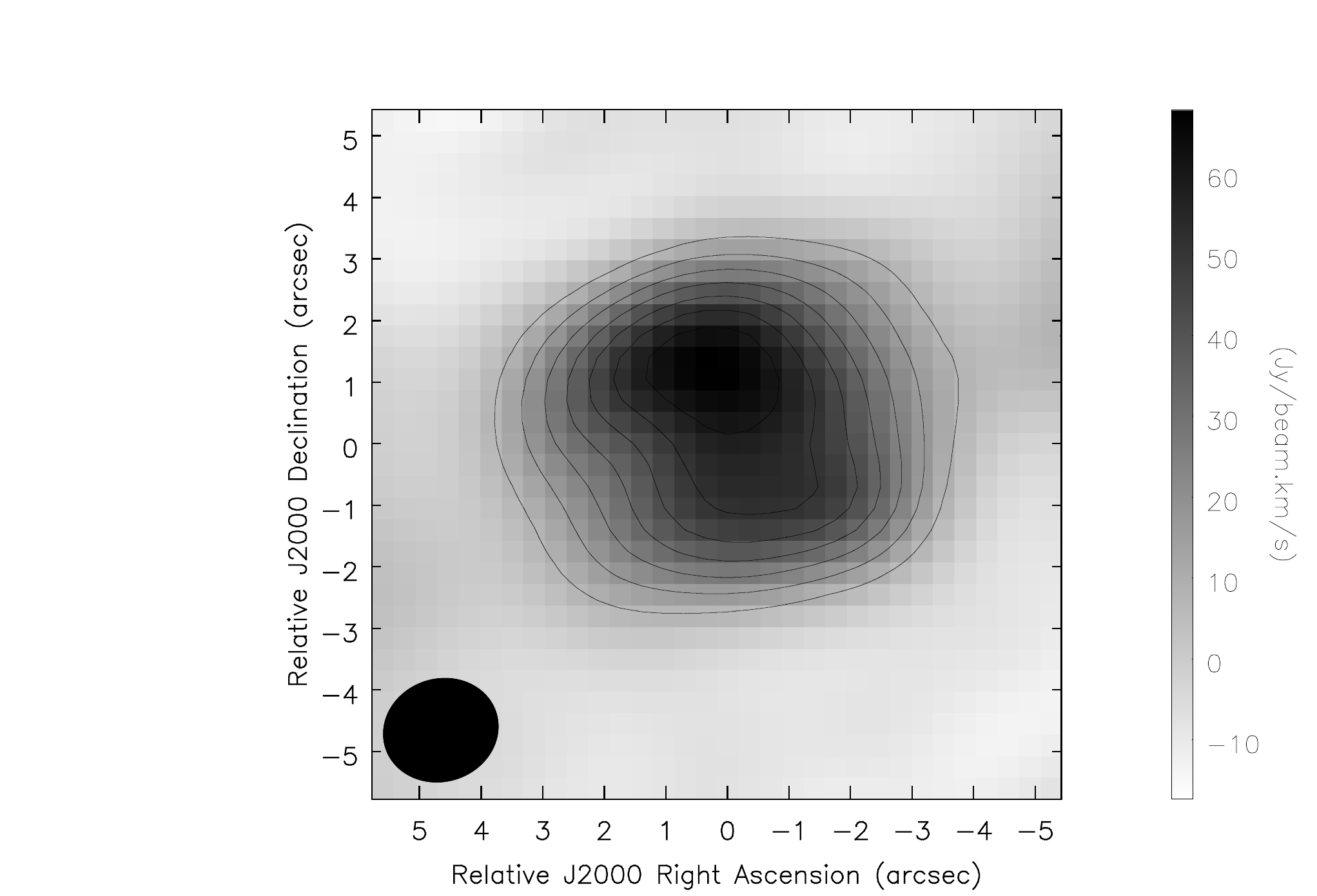}
\caption{Contour 0-moment map of the CO $J=3\rightarrow$2 emission of Frosty Leo. The contours are drawn in steps of $10\times$(1,2,3,4,5) percent of the peak 
and the wedge on the right indicates the continuum flux in Jy\,beam$^{-1}$\,km\,s$^{-1}$. The beam is drawn in the left-bottom corner of each panel.}
\label{CO1}
\end{figure} 

\begin{figure}
\hspace{-1.5cm}
\includegraphics[height=7cm]{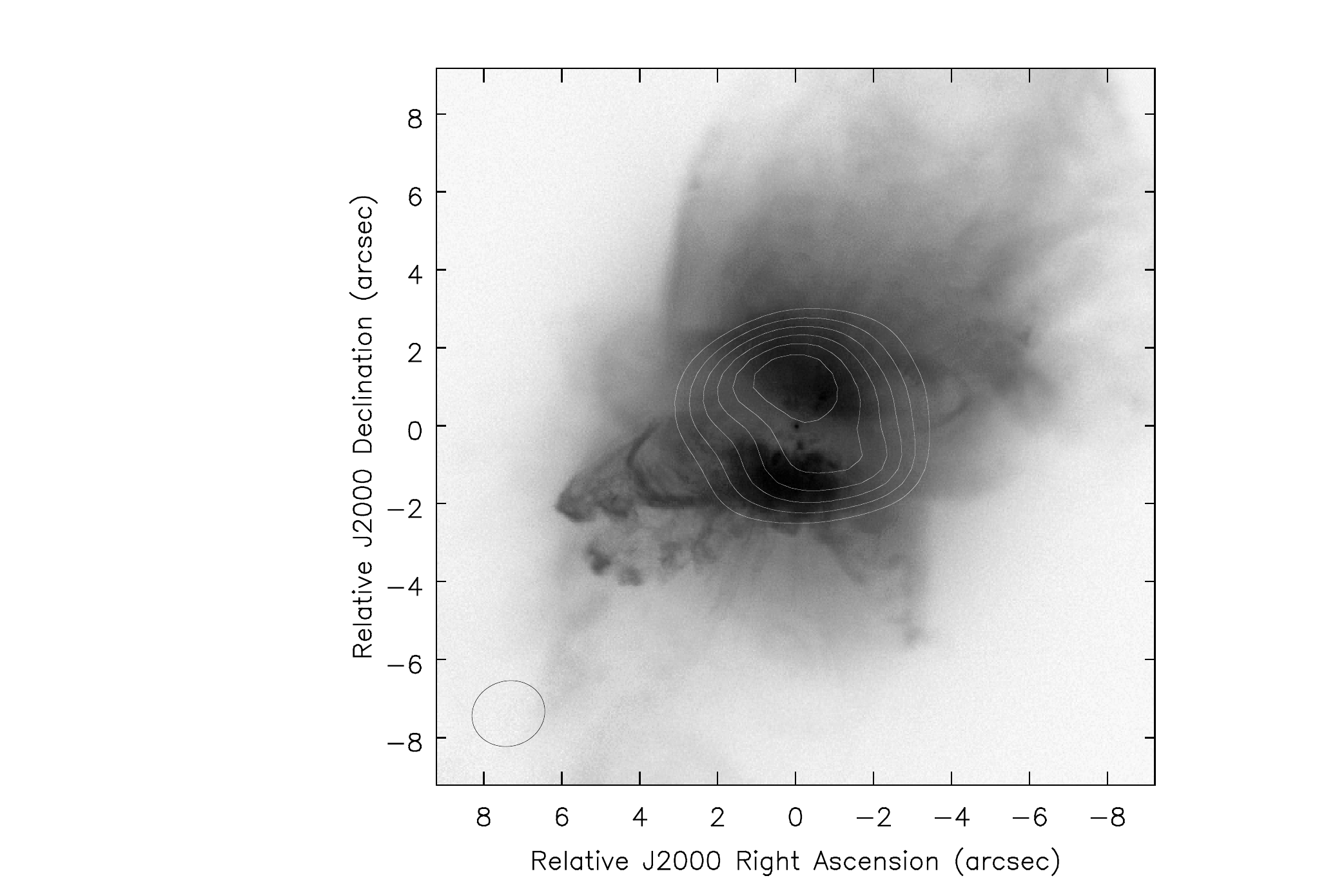}
\caption{HST composite images (F606W and F814W filters) of Frosty Leo with the CO emission contours superimposed. The latter are drawn in steps of 10\% $\times$(3 to 9) of the peak emission.}
\label{hst_CO}
\end{figure} 

\section[]{Dust continuum analysis}\label{continuum}

\begin{figure*}
\begin{center}
\includegraphics[scale=0.7,angle=-90]{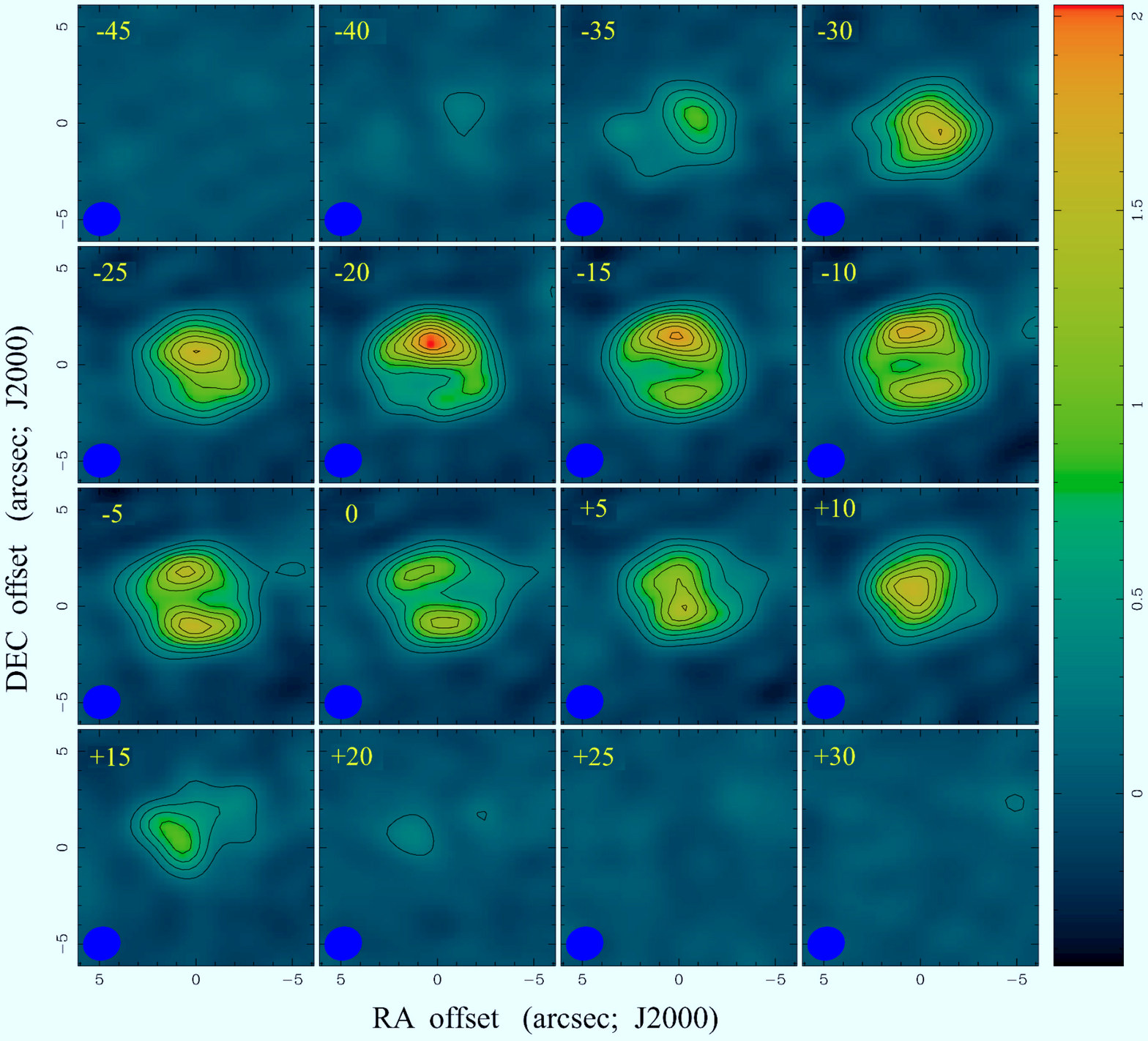}
\caption{Contour maps of the CO $J=3\rightarrow$2 emission of Frosty Leo. The velocities are shown in the left-upper corners. The contours are drawn in steps 10\% $\times$(1 to 9) of the peak emission and the wedge on the right indicates the continuum flux in Jy beam$^{-1}$. The beam is drawn in the left-bottom corner of each panel. 
Coordinates of the origin are $\alpha({\rm J 2000})=09^{\rm h}39^{\rm m}53\fs959$, 
$\delta({\rm J 2000})=+11\degr58\arcmin52\farcs60$. North is up and east is left.}
\label{CO2}
\end{center}
\end{figure*} 

\subsection{Thermal continuum}

In each spectral band the continuum was carefully selected and both LSB and USB datasets were combined. During the data 
reduction process, we used the robust weighting and obtained a synthesized beam with FWHM of 2.20\arcsec$\times$1.85\arcsec and a position angle of 
$-73.9\degr$. The calculated rms noise for the different Stokes images were $\sigma_{I}$=1.67 mJy beam$^{-1}$ and $\sigma_{Q,U}$=1.24 mJy beam$^{-1}$. 

The thermal continuum emission extends over a relatively square area of $\sim$4.9\,arcsec$^{2}$ 
(see Fig.~\ref{cont}). The internal structure displays two bright 
compact clumps centred at $\alpha_{1}$(J2000) = 09$^{\rm h}$ 39$^{\rm m}$ 54\fs000, $\delta_{1}$(J2000) = 
$+11\degr$ 58$\arcmin$ 54\farcs07 and 
$\alpha_{2}$(J2000) = 09$^{\rm h}$ 39$^{\rm m}$ 53\fs893, $\delta_{2}$(J2000) = $+11\degr$ 58$\arcmin$ 
51\farcs95 and with intensity peaks of 
14.2\,mJy\,beam$^{-1}$ (``x1'') and 14.4 mJy beam$^{-1}$ (``x2'') respectively. 
We measured a mean intensity of $\sim$8.1 mJy beam$^{-1}$ over the whole area. 
These values confirm the low continuum emission in Frosty Leo, which are much fainter than those found for 
other PPNe such as CRL 618 and OH\,231.8+4.2 
(see \citealt{Sabin2014}).

Fig.~\ref{hst} shows the distribution of the SMA map compared to observations performed with the {\it Hubble 
Space Telescope}. The figure presents an optical composite image 
of Frosty Leo using F606W filter ($\lambda_{c}= 5997$\AA, $\Delta\lambda$=1502\AA) taken with the {\it 
Wide Field Planetary Camera 2} (WFPC2) as part of the program ID:6816 
(P.I.: R. Sahai), and F814W filter ($\lambda_{c}$=8115.4\AA, $\Delta\lambda$=702.4\AA) taken with the 
Advanced Camera for Surveys (ACS), as part of the program ID:9463 (P.I.: R. Sahai).

The submillimeter continuum emission is located in the central region of the PPN. The line joining the 
two submillimeter emission peaks is not fully coincident with any axes of the optical emission. For instance, 
we have estimated a deviation of $23\degr\pm1\degr$ between such line and the major axis of the PPN.

\subsection{Polarization}

The linear polarization analysis shows that the main polarized emission is extended over an elongated area of $\sim$ 3.1\arcsec$\times$1.8\arcsec. Due to a beam size of 
2.20\arcsec$\times$1.85\arcsec, the ''polarized'' region is therefore barely spatially resolved (in Fig.~\ref{sma} we therefore only show a single polarization segment). 
For the same reason, the smaller polarized spot located in the southern area of the continuum emission (Stokes {\it I}) can be discarded as it is not a meaningful structure.

We measured a peak polarization of $\sim$3.2 mJy beam$^{-1}$ ($\sim$2.6$\sigma$) and a mean of $\sim$2.7 mJy beam$^{-1}$ over the polarized area. 
Assuming a minimum value of 3$\sigma$ for a detection to be considered as more robust, we can only infer that, {\it if real}, a marginal continuum polarization is present in Frosty Leo. Hence, all the results obtained relative to the percentage polarization 
will show an upper limit of $33\%\pm13\%$ (measured at the peak percentage polarization), and those related to the position angles (P.A.) have errors from 12 to 17\degr.
If we compare the distribution and maximum value of the polarized emission from Frosty Leo with that of other PPN studied with SMA (namely, CRL 618 and OH\,231.8+4.2), we observe that Frosty Leo seems to mirror the polarization pattern of OH\,231.8+4.2 in the sense that it is not located in the central region of the nebula, but mostly
on the edge. In addition, the peak polarization intensity of Frosty Leo, has a smaller value compared to the PPNe aforementioned (3 and 5 times smaller, respectively), but this result is expected given the small value of the total  continuum flux intensity (Stokes $I$).\\
The peak polarization and the peak continuum emission do not coincide spatially. Such behaviour is not unusual and can be due to the difference either in opacity, grain size or grain distribution  (see \citealt{Hull2017,Hildebrand1999}), the inefficiency of grain alignment but also due to geometrical effects i.e. the projection of magnetic fields \citep{Frau2011}. \citet{Goncalves2005} show in their figure 6 how the degree of polarization varies as function of the peak intensity at 850 $\mu$m for different inclination angles.\\

With only a marginal detection one has to be cautious regarding the interpretation of the results, particularly 
when dealing with the electric field polarization ($\bmath{E}$-field) and the related magnetic field orientation, based on the theory of dust alignment (\citealt{Lazarian2003,Lazarian2011}).
While we assumed that the dust polarization is connected to the presence of magnetic fields, another process has been recently invoked to explain the polarization of dust grains: self-scattering from randomly aligned dust particles (\citealt{Kataoka2015,Yang2016}). In the case of Frosty Leo, the data in hand do not allow us to clearly separate the two processes, although the polarization vector distribution appears to indicate that dust scattering is not the main process at work. Higher resolution and multi-wavelength data are required for a more definitive assessment of the polarized emission.

\section[]{Molecular line analysis}\label{line}

\begin{figure*}
\hspace{-0.3cm}
\includegraphics[scale=0.6]{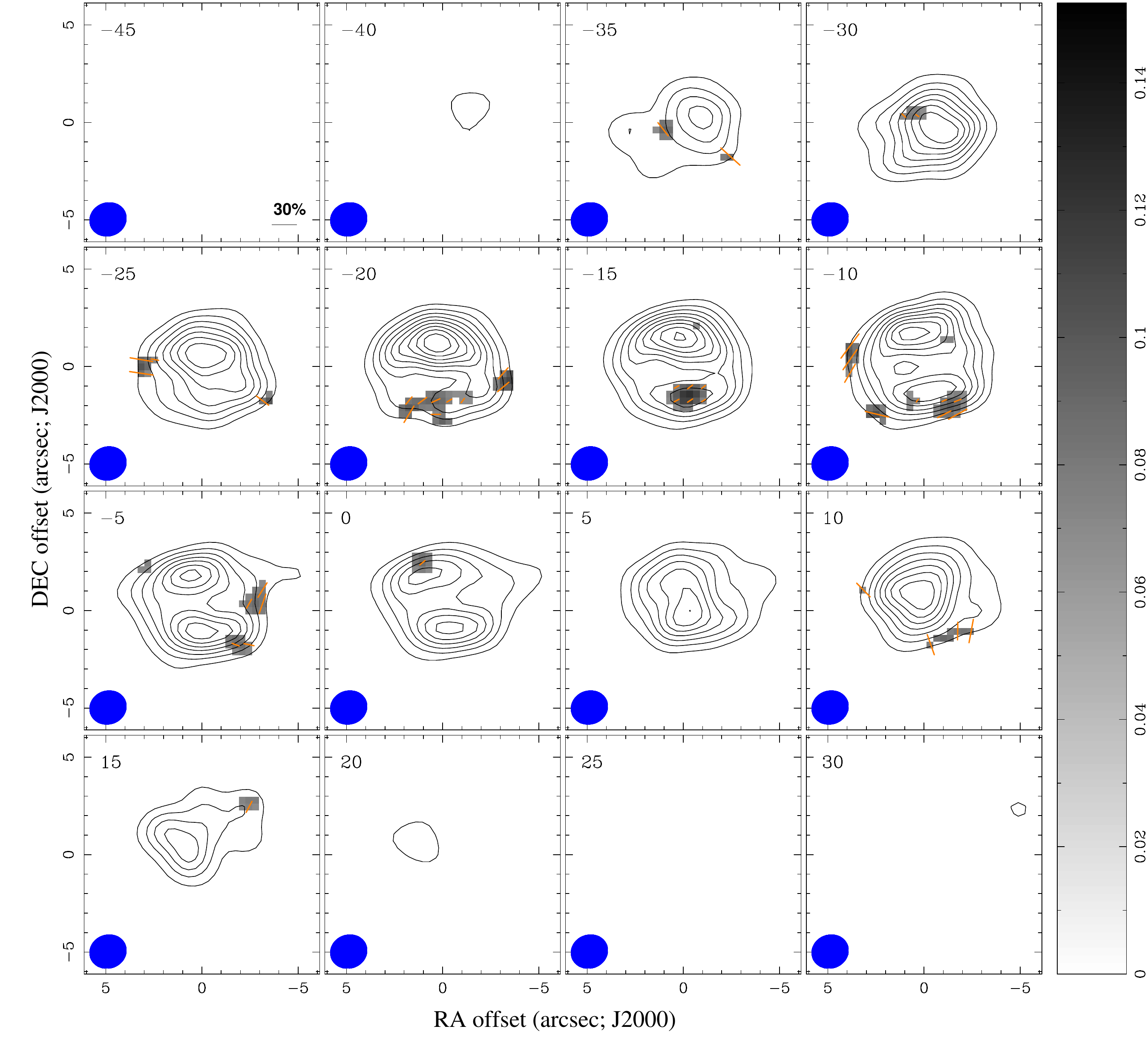}
\caption{Polarization maps of the CO $J=3\rightarrow$2.  The contours correspond to the full intensity while the grey scale identifies the location of the polarized 
emission with a 2.5$\sigma$ cut. The orange bars indicate the E-segments distribution. The emission channels are averaged across intervals of 5 km\,s$^{-1}$ width 
and correspond to central velocities $-21$ and $-15$\,km\,s$^{-1}$. The contours are drawn in steps of 10 $\times$ (1 to 9) percent of the peak and the wedge on the 
right indicates the polarized flux in Jy Beam$^{-1}$. The beam is drawn in the left-bottom corner of each panel. 
Coordinates of the origin are $\alpha({\rm J 2000})=09^{\rm h}39^{\rm m}53\fs959$, 
$\delta({\rm J 2000})=+11\degr58\arcmin52\farcs60$. North is up and east is left.
The scale bar in the right-bottom corner in the first panel ($-45$\,km\,s$^{-1}$) represents the percentage of 
polarization and applies to all the maps (see text).}
\label{CO3}
\end{figure*}


\subsection{Kinematics}

The only molecular line clearly detected in the spectrum of Frosty Leo, in the range 334--346 GHz, is the CO 
$J=3\rightarrow$2. We measured an rms noise $\sigma_{I}= 75.4$ mJy beam$^{-1}$ and $\sigma_{Q,U}$= 33.9 mJy 
beam$^{-1}$  per 5 kms$^{-1}$ channel. Theses values are, as expected, much larger that the continuum ones. The 
FWHM gaussian synthesized beam obtained from the robust weighting was 1.89$\times$1.69 arcsec$^{2}$ with 
PA=$-72.3\degr$.

The analysis of the double peak emission line indicated a systemic velocity $V_{\rm LSR}= -10$ km\,s$^{-1}$ for the 
nebula, in agreement with the literature. Fig.~\ref{CO1} shows the full CO emission (0-moment map) which peaks at $
\sim$68.1 Jy beam$^{-1}$ km\,s$^{-1}$ (with a mean of $\sim$39.1 Jy beam$^{-1}$ km\,s$^{-1}$ on the whole area) and 
extends over $\sim$8.1$\times$6.8 arcsec$^{2}$. This distribution is also presented in Fig.~\ref{hst_CO} where we
compare it to an optical HST image (Fig.~\ref{hst}). We observed that the CO peak emission is coincident with the northern
inner bright region of the nebula, as well the northern continuum peak intensity.

\begin{figure*}
\hspace{-0.7cm}
\includegraphics[scale=0.6]{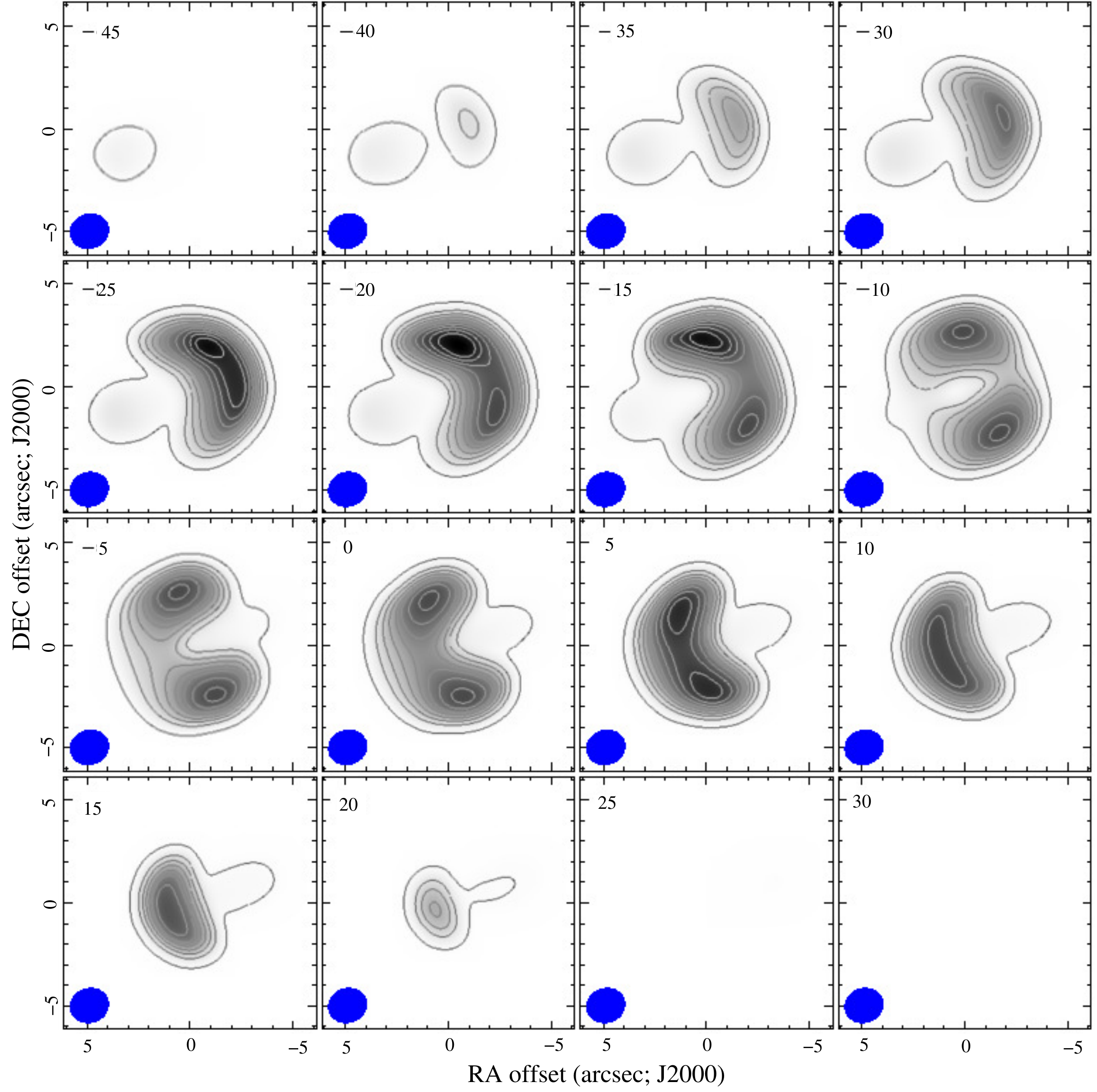}
\caption{Synthetic contour maps of the CO emission obtained with {\sc shape} which globally fit with the observations 
presented in Fig.\ref{CO3}.}
\label{GRID}
\end{figure*}

\begin{figure*}
\hspace{-0.7cm}
\includegraphics[height=7cm]{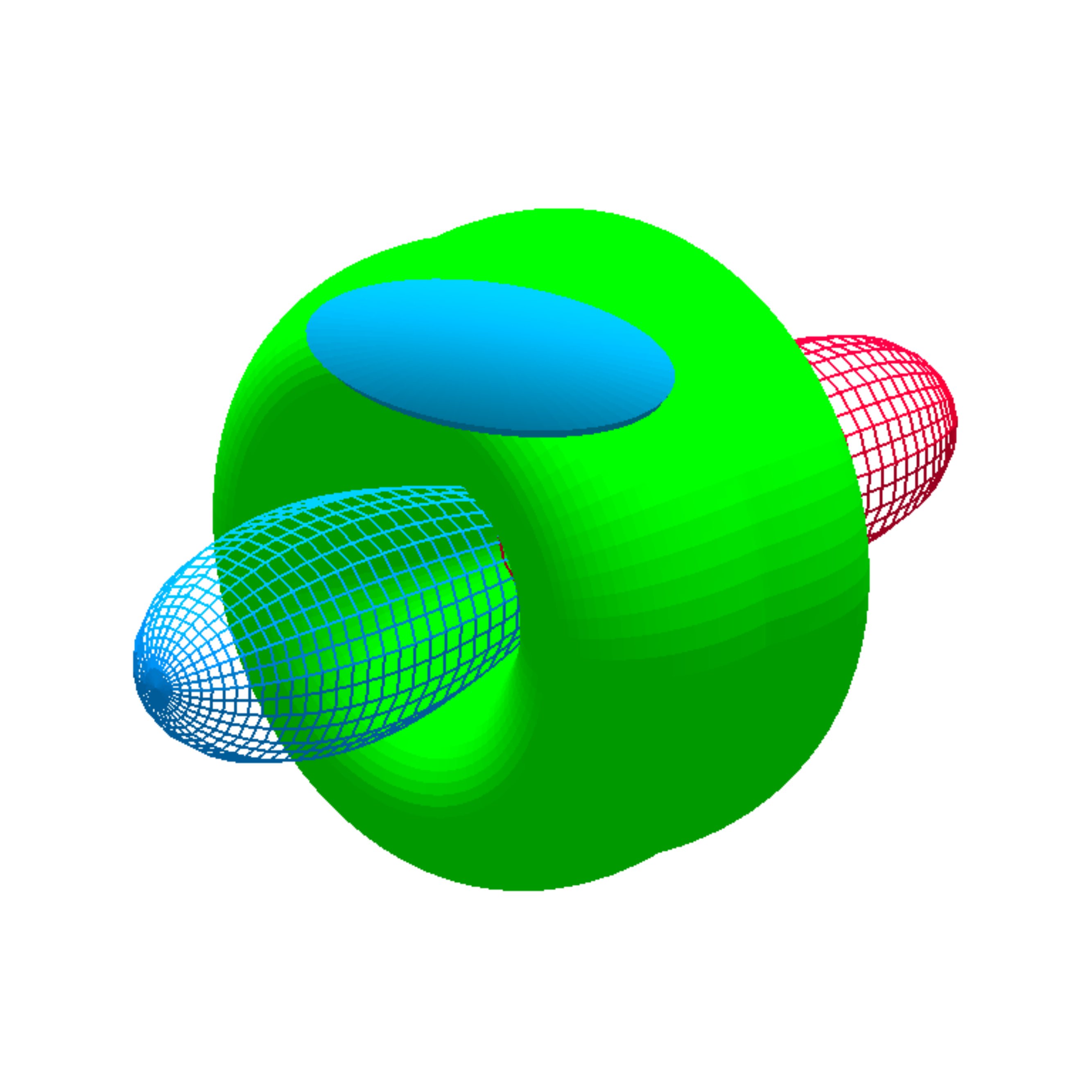}
\includegraphics[height=7cm]{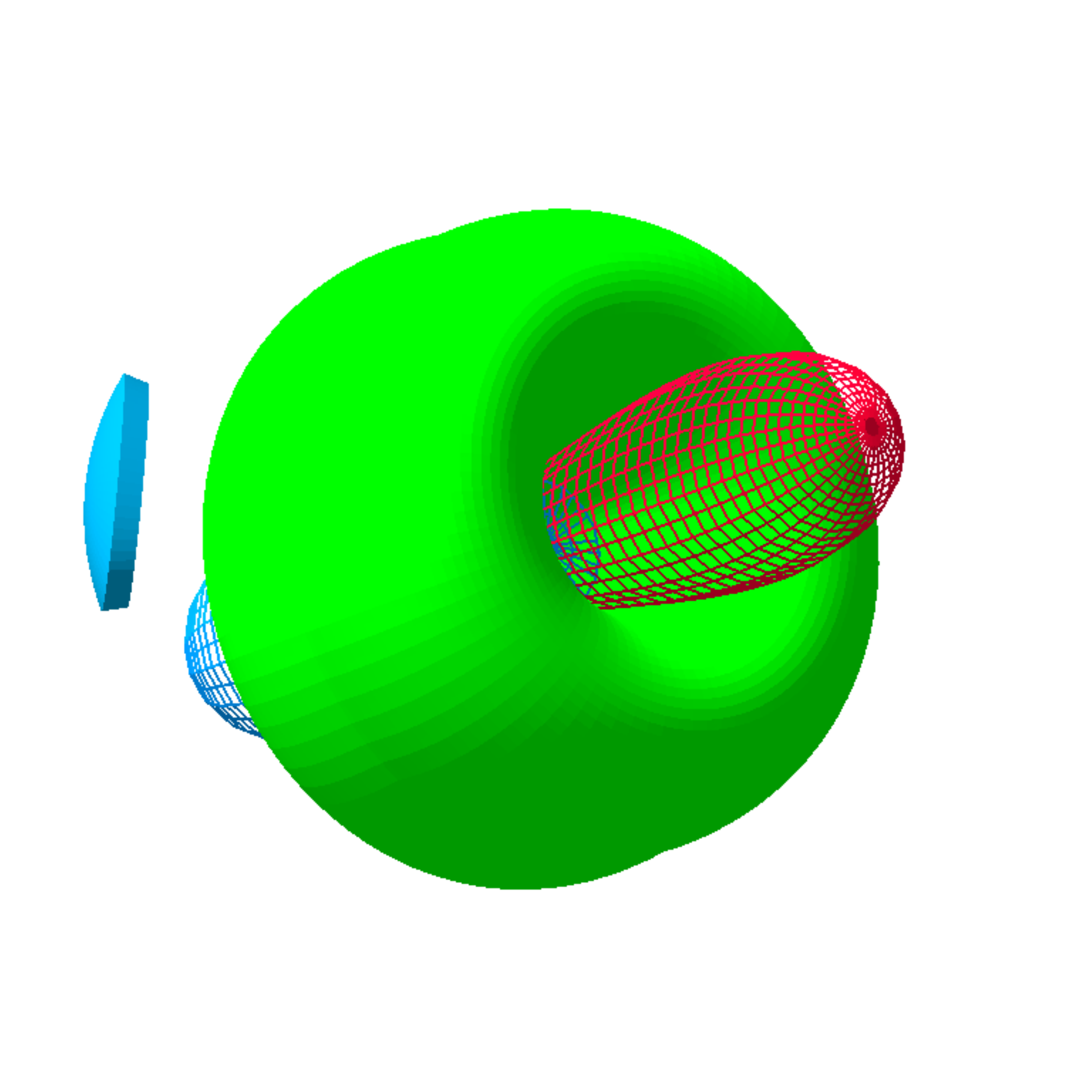}
\caption{Two viewing angle of our {\sc SHAPE} 3D model showing the distribution of the CO $J=3\rightarrow$2 emission in Frosty Leo. 
The observed results are best described by the presence of a distorted torus associated to an emerging outflow and a outside ``cap''.}
\label{MODEL}
\end{figure*}

In Fig.~\ref{CO2} we show contour maps of different velocity channels of the CO $J=3\rightarrow$2 emission of Frosty 
Leo. The maps correspond to a velocity range from $-45$ to +30\,km\,s$^{-1}$, in steps of 5 km\,s$^{-1}$ and a peak 
emission of 2.03 Jy beam$^{-1}$ is observed at $\sim-20$\,km\,s$^{-1}$. 
The change in the shape of the CO molecular line can be observed along the different panels with the more symmetric 
pattern appearing at $-10$\,km\,s$^{-1}$ i.e. the systemic velocity. Our data are compatible with the CO 
$J=2\rightarrow$1 emission observed with the IRAM interferometer by CC05 (their figures 2 and 6), and also with the 
predictions they show in their Fig. 8.

\subsection{Polarization}

As discussed by \citet{Sabin2014}, the Goldreich--Kylafis effect \citep{Goldreich1981,Goldreich1982,Kylafis1983} 
is generally invoked 
when investigating the emission from (rotating) molecules in the presence of a magnetic field. The polarization 
percentage is generally low, on the order of few percents, occurring when anisotropic radiation disturbs the molecular magnetic 
sublevels.


The results of the molecular line polarization analysis conducted for the CO line are presented in Fig.~\ref{CO3}.
Polarization is seen in several channels and we measured a peak of $P_{\rm peak}$=0.13 Jy beam$^{-1}$ corresponding 
to a 3.8$\sigma$ detection at channel $-15$ km\,s$^{-1}$. Similarly to the dust continuum, the molecular polarized 
emission can also be considered a marginal detection assuming that we would expect a minimum of 4 to 5$\sigma$ 
detection to qualify the detection as ``robust''. Hence, the polarization percentage obtained are here again upper limits.
  
However, we observe that, with the velocity interval used, the most internal channels centred at $-15$ and $-20$ km\,s$^{-1}$ indicate the southern 
section of the molecular emission as the location of the strongest polarization intensity (above 3$\sigma$). Also, the overall picture tends to show that the 
mean PA of the corresponding polarization segments is $\sim122\pm11\degr$. This direction seems
correlated with that of an outflow/jet in the PPN within the errors.

\subsection{Implications for the magnetic field}

While the polarization of the thermal dust emission is known to reflect and indicate the presence and direction of 
magnetic fields (via the 90 degrees rotation of the electric field), the same cannot be said for the polarization of 
molecular line. Thus, according to the Goldreich-Kylafis effect, the polarization directions (or position angles) 
are linked to the magnetic field direction on the plane of the sky. Most importantly, the optical depth (and hence the velocity gradient) at which the polarization 
process is occurring is a fundamental parameter as it will determine if the observed polarization is parallel or perpendicular to the magnetic field. In their 
study of the protostellar core NGC\,1333\,IRAS\,4A and its outflows, \citet{Ching2016} measured the  difference in position angle between the magnetic 
segments indicated by the dust polarization analysis and the electric segments obtained by the CO polarization observations (see their figure 4). 
The fundamental assumption being the magnetic field direction is given by the polarized dust emission, they were able to assess the direction of the CO polarization lines with respect to the field.

Thus, we compared the CO polarization maps in Fig. \ref{CO3} with the dust polarization map rotated by 
90$\degr$ in Fig. \ref{sma}. We focused on the brightest regions of the polarized molecular emission and as a result 
we measured a mean P. A. of $122\pm11\degr$ with respect to the channel centred at $-15$\,km\,s$^{-1}$. 
The mean P. A. of the magnetic vector is $96\pm13\degr$.
Therefore, the mean deviation between both datasets is $\sim26\pm17\degr$.
In this case, due to the combined uncertainties in the P.A. of all the parameters involved, as well as the size of the beam, we conclude that it is not possible to infer the direction of the CO polarized emission with respect to
the magnetic field (e.g. parallel or perpendicular).

\section[]{CO Modelling}\label{model}

{\sc shape} \citep{Steffen2011} is an interactive morpho-kinematic modelling software which allows to choose a 
structure from its catalogue (spheres, tori, cones, etc.), and change location, size, and orientation of such structure, among many other variables, including the velocity law and density. Generally, the {\sc shape} user  proposes a structure and interactively moves its parameters (graphically), until the morphology and kinematics from the model are in good agreement with observations. New structures can be added to complete the model. The higher spectral and spatial resolution, the better fit can be achieved.

In order to gain a better insight of the molecular system and the relationship with the marginal polarization emission, 
we have used {\sc shape} to build  a `toy-model' of Frosty Leo based on the observed CO data. Such a model would 
therefore correspond to and describe the observed channel maps mosaic. 
The resulting channel map mosaic (Fig.~\ref{GRID}), presents the following features: (a) we identified two maxima of 
the emission, better seen in the centre channels. This double-peak feature tends to become one peak in the most 
extreme channels, both blueshifted and redshifted and (b) in addition, a marginal evidence of a collimated bipolar 
outflow can be traced.

To model these features using {\sc shape}, we have proposed the existence of three morphological structures: 
an expansive cylindrical ring, a bipolar outflow, and finally a dense small region (hereafter referred to as ``cap''), 
which is separated by {5\,{\arcsec} from the central star towards the North, de-projected, and misaligned 
with respect to the main axis.

Thus, our final model for CO $J=3\rightarrow$2 corresponds to a structure formed by an asymmetrically 
distorted torus and a collimated outflow emerging from the centre (Fig.~\ref{MODEL}). The axis of the torus is aligned 
with the outflow, similarly to the proposition by CC05 in their Fig.~7 for CO $J=2\rightarrow$1 and $J=1\rightarrow$0 
(the so called jets). The torus has a size from 3 to 5.8\,arcsec in diameter (the inner and outer size respectively), 
and an expansion velocity around 20 km\,s$^{-1}$.  The position angle of the torus axis is PA=$110\degr$, and its 
inclination angle is $i=40\degr$. Distortions in the structure of the torus, must be caused by a non-uniform
development of the mass-loss process.

It is important to note that, we have measured the same systemic velocity as CC05, as well as estimated a similar expansion velocity for the torus, and also detected the presence of an outflow (jet) aligned to the torus axis.
The small structure called the ``cap'' was introduced to reproduce the enhanced emission in the surroundings of the North peak, inferred from the contour maps in channels from $-20$ to $-10$\,km\,s$^{-1}$ (Fig.~6). This structure could be an isolated off-axis ejection from the nucleus of the nebula, but it is hard to explain the mechanism that produces such ejection. Another possibility is that this ``cap'' is a clump of gas from the interstellar medium, with an enhanced density with respect to its environment.  Such structure could be excited by the interaction with the stellar mass loss from the Frosty Leo's nucleus. Certainly, we need a higher spatial and spectral resolution to figure out the real nature of this structure.\\

Thanks to this qualitative model, it is now possible to disentangle the different components of the CO emission and therefore associate them to the polarized emission. When comparing Fig.~\ref{CO3} and Fig.~\ref{GRID} we observe that the CO polarization is mostly linked to the torus and marginally to the outflow/jet.

\section[]{Conclusions}\label{conclusion}

We present the SMA polarization observations of Frosty Leo.
We measured peaks of 14.4\,mJy beam$^{-1}$ and 68.1 Jy beam$^{-1}$ km\,s$^{-1}$ for the dust continuum and the CO 
$J=3\rightarrow$2 emission in Stokes {\it I} respectively. The polarized emission detection, in both cases, can be considered ``marginal'' 
 with peaks at 2.6$\sigma$ (dust) and 3.8$\sigma$ (CO). Therefore the subsequent results have to be taken with caution. 
 
In the case of the dust continuum, if the polarization information is real, the $\bmath{B}$-segments would likely be parallel to the equatorial plane of 
Frosty Leo and would therefore trace (part of) a toroidal magnetic field. The emission from CO $J=3\rightarrow$2 is in total agreement with those 
detected by \citep{Carrizo2005} in CO $J=2\rightarrow$1. 
The comparison of the molecular channel polarization maps with the dust polarization map (rotated by 90\degr) indicates that the $\bmath{E}$-segments associated to the most polarized area displays a $\sim$23\degr difference in position angle with respect to the $\bmath{B}$-segment derived from the dust polarization. A conclusion would be that the $\bmath{E}$-segments are likely parallel to the magnetic field in this area. But again, the uncertainty of the PA has to be acknowledged.

Finally, using the kinematical information from the CO $J=3\rightarrow$2 emission, we were able to generate a toy model indicating the various 
components at play. We identified a distorted torus, a bipolar outflow or jet aligned with the torus' axis and a flattened spherical ``cap''. The comparison 
of the location of the CO polarization segments with our model suggests that the polarized emission are mostly linked to the torus and in a lesser extent 
to the jet/outflow. 

The low detection levels are preventing us from drawing a clearer picture of the polarization state of Frosty Leo and making more speculation on the 
magnetic field structure; but the observations presented here give us a starting point which would greatly benefit from deeper (larger collecting area, 
higher resolution) polarimetric observations. 
 


\section*{Acknowledgments}

The authors would like to thank the referee for carefully reading our manuscript and for her/his comments which helped improving the quality of the article. We also thank the SMA staff for supporting the observations. LS acknowledges support from PAPIIT grant IA-101316 (Mexico). The Submillimeter Array is a joint project between the Smithsonian Astrophysical Observatory and the Academia Sinica Institute of Astronomy and Astrophysics and is funded by the Smithsonian Institution and the Academia Sinica. Some of the 
data presented in this paper were obtained from the Mikulski Archive for Space Telescopes (MAST). STScI is operated by the Association of Universities 
for Research in Astronomy, Inc., under NASA contract NAS5-26555. Support for MAST for non-HST data is provided by the NASA Office of Space Science 
via grant NNX09AF08G and by other grants and contracts.

\bibliographystyle{mn2e}

\bibliography{sabin_FL}

\bsp


\end{document}